\documentclass[11pt,a4paper]{article}

\pdfoutput=1
  
\usepackage{amsmath,amssymb}
\usepackage{authblk}
\usepackage[export]{adjustbox}

\usepackage{epsfig}  
\usepackage{graphicx}               
\usepackage{url}
\usepackage[colorlinks=true,linkcolor=blue,citecolor=blue,urlcolor=blue]{hyperref}
\usepackage{cite}
\usepackage{caption}
\usepackage{subcaption}

\usepackage{color}

\usepackage{microtype}

\usepackage[margin=2.5cm]{geometry}

\title{  
\vspace*{-2.3cm}  
\begin{flushright}  
{\normalsize{ 
CERN-PH-TH-2015-093
  }  }
\end{flushright}  
\vspace{2.5cm}  
\Huge  
\textbf{
Gravitational Waves From a Dark (Twin) Phase Transition
}\vspace*{0.5cm}   
}

\author{\bf Pedro Schwaller\thanks{pedro.schwaller@cern.ch}}
\affil{ CERN, Theory Division, CH-1211 Geneva 23, Switzerland}

\date{\today}

\begin{document}

%
\maketitle  
\begin{abstract} 
In this work, we show that a large class of models with a composite dark sector undergo a strong first order phase transition in the early universe, which could lead to a detectable gravitational wave signal. We summarise the basic conditions for a strong first order phase transition for SU(N) dark sectors with $n_f$ flavours, calculate the gravitational wave spectrum and show that, depending on the dark confinement scale, it can be detected at eLISA or in pulsar timing array experiments. The gravitational wave signal provides a unique test of the gravitational interactions of a dark sector, and we discuss the complementarity with conventional searches for new dark sectors. The discussion includes Twin Higgs and SIMP models as well as symmetric and asymmetric composite dark matter scenarios. 
\end{abstract}  
%
%

  


 
  

\section{Introduction}
\label{sec:intro}

Violent phenomena in the early universe can lead to large anisotropic fluctuations in the energy momentum tensor, which can act as sources for gravitational waves (GW). Strong first order phase transitions (PT) are an example of such a phenomenon, and it is well known that they can produce GWs~\cite{Witten:1984rs,Hogan:1984hx,Hogan:1986qda,Turner:1990rc}. 
Once produced, GWs propagate through space almost undisturbed, and can therefore serve as a unique probe of phenomena in the early universe. 

Phase transitions in particle physics are usually associated with symmetry breaking, i.e. with the transition of the universe from a symmetric phase to a phase of broken symmetry, as the temperature drops below a critical temperature $T_*$. Within the standard model (SM) of particle physics, at least two phase transitions should take place, associated with the breaking of electroweak symmetry around $T_*\sim100$~GeV and with the breaking of chiral symmetry at the time of the QCD phase transition, $T_* \sim 0.1$~GeV. 

Today we know that both the QCD and the electroweak phase transition are not first order, but proceed through a smooth cross-over~\cite{Aoki:2006we,Bhattacharya:2014ara,Kajantie:1995kf,Kajantie:1996mn}, and can therefore not produce a strong GW signal through the usual mechanism (see however~\cite{Ghiglieri:2015nfa}). This can however be changed in models beyond the SM. Extensions of the SM which lead to a strong first order electroweak phase transition are particularly attractive since they can provide one missing ingredient for generating the observed baryon asymmetry of the universe. GW signals from such models have for example been studied in~\cite{Grojean:2006bp,Delaunay:2007wb,Huber:2007vva,Das:2009ue,Espinosa:2008kw,Jarvinen:2009mh,Espinosa:2010hh}. It is more difficult to modify the QCD phase transition, although a large neutrino chemical potential could be sufficient to provide a strong first order PT~\cite{Schwarz:2009ii}. The resulting signal was studied in~\cite{Caprini:2010xv}. 

The aim of this work is to point out that gravitational waves could also be produced by a strong PT in a dark or hidden sector. The particular scenario we have in mind is a dark sector with a new SU($N_d$) gauge interaction which confines at some scale $\Lambda_d$. Such models have recently received renewed interest either as models of dark matter~\cite{Kribs:2009fy,Alves:2009nf,Falkowski:2009yz,Blennow:2010qp,Frandsen:2011kt,Feng:2011ik,Kumar:2011iy,Buckley:2012ky,Hambye:2013dgv,Bai:2013xga,Bhattacharya:2013kma,Cline:2013zca,Boddy:2014yra,Hur:2011sv,Newstead:2014jva,Krnjaic:2014xza,Detmold:2014qqa,Heikinheimo:2014xza,Yamanaka:2014pva,Dorsch:2014qpa,Appelquist:2013ms,Appelquist:2015yfa,Schwaller:2015gea,Cohen:2015toa,Antipin:2015xia,Carmona:2015haa} or as part of the low energy sector of so called Twin Higgs models~\cite{Chacko:2005pe,Chacko:2005un,Craig:2013fga,Craig:2015pha,Barbieri:2015lqa,Low:2015nqa,Batell:2015aha}. Different from generic hidden sectors~\cite{Strassler:2006im}, these models provide a preferred mass range and some restrictions on the particle content, such that the frequency range of the potential GW signal can be predicted. 

Given that the SM QCD transition is not first order, we will review the known results on the order of the PT in strongly coupled gauge theories in the next section, followed by a discussion of models that fall into this category. In Sec.~\ref{sec:gws} we calculate the GW spectra that can be produced in these models, and compare them to the sensitivity of current and planned GW detection experiments in Sec.~\ref{sec:detect}. We discuss the complementarity of GW experiments with other searches for dark sectors in Sec.~\ref{sec:compl}, before presenting our conclusions. 

%
\section{Models with First Order Phase Transition}
\label{sec:models}
Near the QCD confinement scale $\Lambda_{\rm QCD}$, the dynamics of QCD is governed by three flavours, two of which are almost massless, while the strange quark mass is of order $\Lambda_{\rm QCD}$. Lattice studies~\cite{Aoki:2006we,Aoki:2006br,Bhattacharya:2014ara} have shown that for these values of the quark masses, the QCD PT is a weak cross-over.

\begin{figure}
\centering
\includegraphics[width=0.45\textwidth]{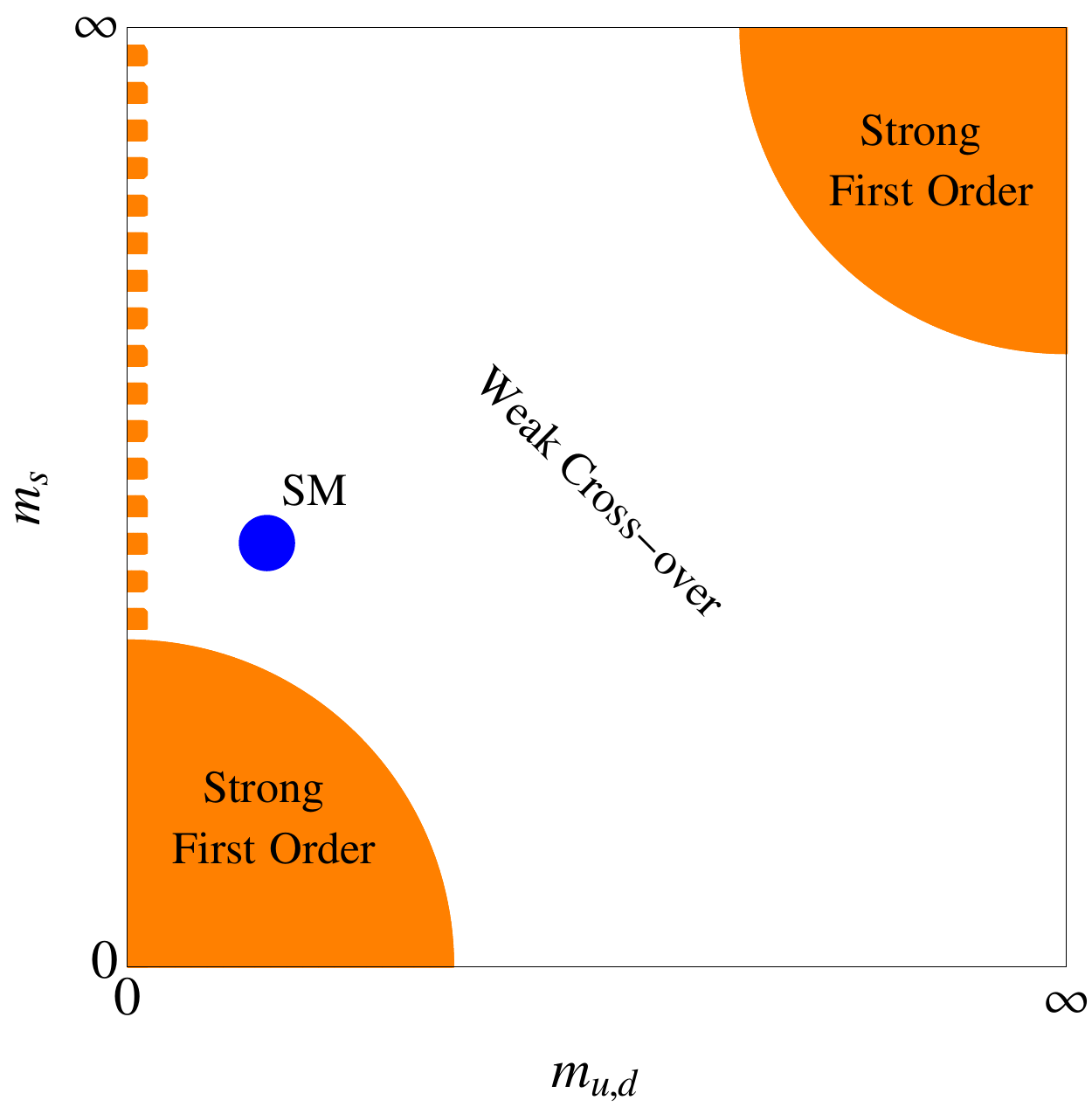}
\caption{Phase diagram of QCD at zero chemical potential (schematic). The dashed region represents our current lack of knowledge about the order of the PT in the limit of two massless flavours. }
\label{fig:columbia}
\end{figure}

However this is not a generic result for QCD and similar theories, but more a consequence of the precise values of $m_u \approx m_d$ and $m_s$ in the SM. The QCD phase diagram for arbitrary $m_{u,d}$ and $m_s$ can be summarised in the so called Columbia plot, which is reproduced in Fig.~\ref{fig:columbia}, based on~\cite{Bonati:2012pe}. The pure Yang-Mills limit $m_{u,d}, m_s \to \infty$ is known to have a strong first order PT~\cite{Svetitsky:1982gs} from the restoration of a global $Z_3$ center symmetry at low temperatures. The opposite $m_{u,d}, m_s \to 0$ limit, i.e. the theory with three exactly massless quarks, also features a strong first order transition, related to the breakdown of the $SU(3)\times SU(3)$ chiral symmetry~\cite{Pisarski:1983ms}. 

Here we are interested in $SU(N_d)$ theories with $N_d \geq 3$ and $n_f$ massless\footnote{More precisely, $n_f$ is the number of fermions with $m \ll \Lambda_d$.} Dirac fermions in the fundamental representation, and with a confinement scale $\Lambda_d$. To guarantee the existence of a confining phase we further impose $n_f < 4N_d$, in order to stay outside of the conformal window. 

For pure Yang-Mills theories, $n_f = 0$, the confinement phase transition is related to the restoration of a global $Z_{N_d} \subset SU(N_d)$ center symmetry, which is broken in the high temperature phase. Lattice simulations have shown that this PT is strong first order for $N_d \geq 3$~\cite{Panero:2009tv}. For the $N_d=3$ case there is also the classic argument of~\cite{Svetitsky:1982gs}: In the case of a second order PT, the critical behaviour of the theory should be described by a $Z_{3}$ symmetric effective theory at an infrared fixed point. Since no fixed points are known for theories with $Z_3$ symmetry, the PT must be first order. 

The case of non-zero $n_f$ for $N_d=3$ and $N_d \to \infty$ is discussed in~\cite{Pisarski:1983ms}. For $n_f=1$ there is no symmetry breaking, and therefore no phase transition. For $n_f \geq 3$ the PT is first order\footnote{The behaviour can be understood by studying the effective theory near the confinement scale, which consists of $n_f^2$ massless Goldstone bosons from the breaking of the $SU(n_f)\times SU(n_f)\times U(1)_A$ chiral symmetry. For $N_d \to\infty$, the anomaly is negligible, and the PT is first order due to the absence of infrared fixed points in the effective theory, provided that $n_f \geq \sqrt{3}$. For  finite $N_d$ and $n_f=3$ the anomaly contribution to the effective lagrangian (i.e. the $\eta'$ mass) is cubic in the Goldstone fields, and therefore alone sufficient to render the PT first order. For $n_f>3$ the anomaly does not affect the PT behaviour, so the large $N_d$ result applies and the PT is first order. What remains unresolved is the $n_f =2$ case for finite $N_d$. For more details the reader is referred to~\cite{Pisarski:1983ms}.} 
for any value of $N_d \geq 3$. In the large $N_d$ limit, the PT is also first order for $n_f=2$.
The only case that is currently not understood is $n_f=2$ for finite $N_d$, where no analytic arguments can be applied and lattice results are difficult to obtain. 

To summarise, $SU(N_d \geq 3)$ theories with $n_f$ massless flavours have a first order PT if either $n_f =0$ or $3 \leq n_f < 4N_d$. 

\

The above discussion hopefully convinced the reader that strong PTs are generic in extensions of the SM which feature new confining gauge symmetries. In the following we will discuss a few examples which are well motivated either from a dark matter perspective or by naturalness arguments. These physically motivated scenarios furthermore provide a preferred range for the confinement scale $\Lambda_d$, which allows us to estimate the temperature of the PT, $T_* \sim \Lambda_d$, and therefore make a prediction for the GW spectrum. 

\paragraph{Composite Dark Matter I (CDM1)} In this class of models, the dark matter candidate is the lightest baryonic bound state of a $SU(N_d)$ dark sector with $n_f$ dark quarks which are neutral with respect to all SM interactions. This allows the DM to be light, since it will only communicate with the SM through heavy mediators. The most natural realisation of these models is in the context of asymmetric dark matter, where the DM number density is related to the baryon asymmetry (see e.g.~\cite{Petraki:2013wwa,Zurek:2013wia,Boucenna:2013wba} for reviews). The measured DM density then implies a mass range of ${\cal O}(5)$~GeV for the DM particle, and therefore motivates $T_* \approx \Lambda_d\sim (1-10)$~GeV. 

As a concrete benchmark we will consider the dark QCD model~\cite{Bai:2013xga,Schwaller:2015gea}, which, as the name suggests, consists of a $SU(N_d=3)$ theory with $n_f \geq 3$ dark quarks, such that a strong PT is guaranteed. 

\paragraph{Composite Dark Matter II (CDM2)} Similar to the previous case, the DM is a baryonic bound state of a new $SU(N_d)$ theory. The important difference now is that the dark quarks carry electroweak quantum numbers. Models of these type were considered for example in~\cite{Appelquist:2013ms,Appelquist:2015yfa}, where mass spectra and form factors are calculated from first principles on the lattice. Constraints on the invisible width of the $Z$-boson immediately suggest a confinement scale at or above the $100$~GeV scale, while unitarity arguments on DM freeze-out place a limit of $\Lambda_d \lesssim {\cal O}(100)$~TeV. For fermionic DM (i.e. for odd $N_d$), direct detection furthermore imposes a bound of $M_{\rm DM} \sim \Lambda_d \gtrsim10$~TeV~\cite{Appelquist:2013ms}. For scalar DM (even $N_d$) lower DM masses are allowed~\cite{Appelquist:2015yfa}. The benchmark model considered in~\cite{Appelquist:2015yfa} has $N_d=4$ with $n_f=4$ flavours and should therefore feature a strong first order PT.\footnote{In~\cite{Appelquist:2015yfa} the quark masses were taken close to the confinement scale to allow a faster simulation. In this case the general arguments for the order of the PT do not apply, and a lattice simulation might be necessary to determine the order of the PT. However the model is also viable phenomenologically with lighter quark masses.}

\paragraph{Twin Higgs (TH)} These models attempt to solve the hierarchy problem without introducing coloured partners for the top quark~\cite{Chacko:2005pe,Chacko:2005un}, and have recently received renewed attention~\cite{Craig:2015pha,Barbieri:2015lqa,Low:2015nqa}. This is achieved by adding a twin sector to the SM with an approximate $Z_2$ parity symmetry, with the minimal requirement that there should be fermionic partners for the top and bottom quarks which are charged under a new $SU(N_d=3)$ interaction. The approximate $Z_2$ symmetry constrains $\Lambda_d$ to lie in the $(1-10)$~GeV range. Furthermore the twin top and bottom partner masses are necessarily much larger than $\Lambda_d$, such that the theory at the PT scale is $SU(3)$ with $n_f=0$, which again predicts a strong first order PT. 

\paragraph{Strongly Interacting Massive Particle (SIMP)} The key idea behind SIMP DM is that the relic density is determined through freeze-out of $3\to 2$ instead of $2\to 2$ annihilations~\cite{Hochberg:2014dra}. The correct relic abundance is then obtained for DM masses in the 100~MeV range. One simple realisation of this mechanism is a $SU(N_d)$ dark sector with dark pions as DM candidates. The $3\to 2$ processes can then be induced by the Wess-Zumino-Witten term~\cite{Hochberg:2014kqa,Lee:2015gsa}. The condition for the existence of the WZW term is $n_f \geq 3$, which again allows for a strong first order PT, this time in the $100~{\rm MeV}-{\rm GeV}$ range. 

\

All these models should give rise to a GW signal from the strong PT, and the resulting spectra and detection possibilities will be discussed in the following sections. Before moving on, we should mention a few other scenarios which also contain new strongly coupled sectors. The first group consists of Hidden Valley~\cite{Strassler:2006im} and vector-like confinement models~\cite{Kilic:2009mi}, which propose the existence of new confining sectors, communicating with the SM either through heavy mediators or directly via SM gauge interactions. If the particle content is such that the models fall into one of the strong PT regions, then also these models will give rise to a GW signal. Hidden sectors also play a role in conformally invariant extensions of the SM~\cite{Heikinheimo:2013fta,Holthausen:2013ota}, and the presence of a strong PT in these models was demonstrated using NJL methods in~\cite{Kubo:2014ida}. 

Furthermore there are so called mirror-world models~\cite{Foot:1991bp}, where the dark sector consists of an exact copy of the SM. Here one does not expect a GW signal for two reasons. First, if the quark content is an exact copy of QCD, there will not be a first order PT. If for some reason the quark masses are modified, such that a strong PT happens, cosmological constraints on the mirror photon require the mirror sector to have a lower temperature. Since the energy density is proportional to the fourth power of the temperature, and the intensity of a GW signal is proportional to the energy density, this will strongly suppress any GW signal from a mirror world.

\section{Gravitational Wave Spectra}\label{sec:gws}

Gravitational waves produced at a time $t_*$ (or equivalently at a temperature $T_*$) will propagate undisturbed in the expanding universe, therefore their frequency $f$ and their fraction of the critical energy density $\Omega_{\rm GW}$ will decrease as $a^{-1}$ and $a^{-4}$, respectively, where $a(t)$ is the scale factor~\cite{Kamionkowski:1993fg,Grojean:2006bp}. Denoting by $a_*$ and $a_0$ the scale factors at time of production and today, entropy conservation ($s a^3 = {\rm const}$) implies 
\begin{align}
	\frac{a_*}{a_0} = \left(\frac{g_{0,s}}{g_{*,s}} \right)^{\frac{1}{3}} \frac{T_0}{T_*} .
\end{align}
Here $T_0 = 2.725\,{\rm K} = 2.348\times 10^{-13}$~GeV is the temperature of the CMB, $g_{*,s}$  ($g_{0,s}=3.91$) is the effective number of relativistic degrees of freedom contributing to the total entropy at the time of production (today), and the entropy density at temperature $T$ is given by 
$s(T) = \frac{(2 \pi)^2}{45} g_s(T) T^3$, where 
\begin{align}
	g_s(T) & = \sum_{i=\rm bosons} g_i \left(\frac{T_i}{T}\right)^{3}+\frac{7}{8}\sum_{i=\rm fermions} g_i \left(\frac{T_i}{T}\right)^{3},
\end{align}
and $g_i$ counts the internal degrees of freedom of the i-th particle. It follows that the frequency today can be expressed as
\begin{align}
	f & = \frac{a_*}{a_0} H_* \frac{f_*}{H_*} = 1.59 \times 10^{-7}~{\rm Hz} \times \left(\frac{g_{*}}{80} \right)^{\frac{1}{6}} \times \left(  \frac{T_*}{1~{\rm GeV}}\right)\times  \frac{f_*}{H_*}\,, 
	\label{eqn:f0}
\end{align}
where we have used the Hubble rate at time of production, $H_* = \sqrt{\frac{4 \pi^3 g_*}{45}} \frac{T_*^2}{M_{\rm Pl}}$, and assumed that all species are in thermal equilibrium at $T=T_*$, i.e. $g_* =g_{*,s}$. For the fraction of energy density in gravitational waves today we similarly obtain 
\begin{align}
	\Omega_{\rm GW} & = \frac{\rho}{\rho_{\rm crit}} = \left( \frac{a_*}{a_0}\right)^4 \frac{H_*^2}{H_0^2}\, \Omega_{*\rm GW}  = 1.77\times 10^{-5} h^{-2} \left( \frac{80}{g_*}\right)^{\frac{1}{3}} \Omega_{*\rm GW} \,,
\end{align}
where we used that $\rho_{\rm crit}/\rho_{*\rm crit} = H_0^2/H_*^2$ and $H_0 = 2.13 \times h \times 10^{-42}$~GeV. It is noteworthy that the intensity of the GW signal is independent of the production temperature $T_*$ (except for the implicit dependence of $g_*$ on $T_*$). 

\

The most sensitive frequency regions of pulsar timing arrays and satellite based experiments are in the nano-Hz and milli-Hz range, respectively. 
To get an idea about the detectability of GWs from a strong dark PT we will therefore need to understand the spectrum of the produced GWs. For this, we will closely follow the discussion of~\cite{Caprini:2010xv}. 
%

Gravitational Waves are sourced by tensor fluctuations of the energy momentum tensor of the primordial plasma. During first order phase transitions both bubble collisions~\cite{Kosowsky:1991ua,Kosowsky:1992vn,Caprini:2007xq} and magnetohydrodynamical (MHD) turbulence~\cite{Hogan:1983zz,Kosowsky:2001xp,Dolgov:2002ra,Caprini:2006jb,Gogoberidze:2007an,Kahniashvili:2008pf,Kahniashvili:2009mf} provide sources of gravitational waves. As functions of the conformal wave number $k = 2 \pi a f$, the GW spectra produced by either source can be approximated by~\cite{Caprini:2010xv}
\begin{align}
	\frac{d\Omega^{(B)}_{\rm GW} h^2}{d \log k} & \simeq \frac{2}{3\pi} h^2 \Omega_{r0} \left(\frac{{\cal H}_*}{\beta}\right)^2 \Omega_{S*}^2 v^3 \frac{(k/\beta)^3}{1+(k/\beta)^4} \,, \label{eqn:bubble} \\
	\frac{d\Omega^{(MHD)}_{\rm GW} h^2}{d \log k} & \simeq \frac{8}{\pi^6} h^2 \Omega_{r0}  \left(\frac{{\cal H}_*}{\beta}\right) \Omega_{S*}^{3/2} v^4 \, \frac{ (k/\beta)^3}{(1+4 k/{\cal H}_*)\left(1+(v/\pi^2)(k/\beta)\right)^{11/3}}\,.\label{eqn:MHD}
\end{align}
Eqn.~(\ref{eqn:bubble}) is based on~\cite{Caprini:2009fx,Huber:2008hg} while Eqn.~(\ref{eqn:MHD}) is a fit to the numerical results of~\cite{Caprini:2009yp}. Here ${\cal H}_*$ is the conformal Hubble parameter ${\cal H} = Ha$ at $T=T_*$, and $\Omega_{r0}$ is the radiation energy density today. The quantities that determine the GW spectrum are the bubble nucleation rate $\beta$ (the duration of the PT is $\beta^{-1}$), the bubble velocity $v$ and the relative energy density in the source, $\Omega_{S*} = \rho_{S*}/\rho_{*,\rm crit} = \Omega_{*\rm GW}$. The temperature of the PT enters through the dependence of ${\cal H}_*$ on $T_*$.

The duration of the PT is usually taken as $(1-100)$\% of a Hubble time, and therefore $\beta = (1-100){\cal H}$~\cite{Hogan:1984hx}. 
To understand the relation with the physical frequency, remember that the conformal frequency is related to the conformal wave number via $a f = k/(2\pi)$. Furthermore using ${\cal H} = Ha$ we see that $f_*/H_* = a f_*/{\cal H}_* = (k/{\cal H}_*)/(2 \pi)$, which together with Eqn.~(\ref{eqn:f0}) allows us to translate the GW spectra into physical frequencies. 

In a given theory, the dynamics of the phase transition, 
and therefore the parameters $\beta$, $v$ and $\Omega_{S*}$, are in principle calculable. For the strongly coupled models considered here they are however not known, and can only be estimated using lattice simulations. We will therefore take $\beta$, $v$ and $\Omega_{S*}$ as additional input parameters, with values motivated by results of analyses in weakly coupled models.

\begin{figure}
\centering
\includegraphics[width=0.45\textwidth]{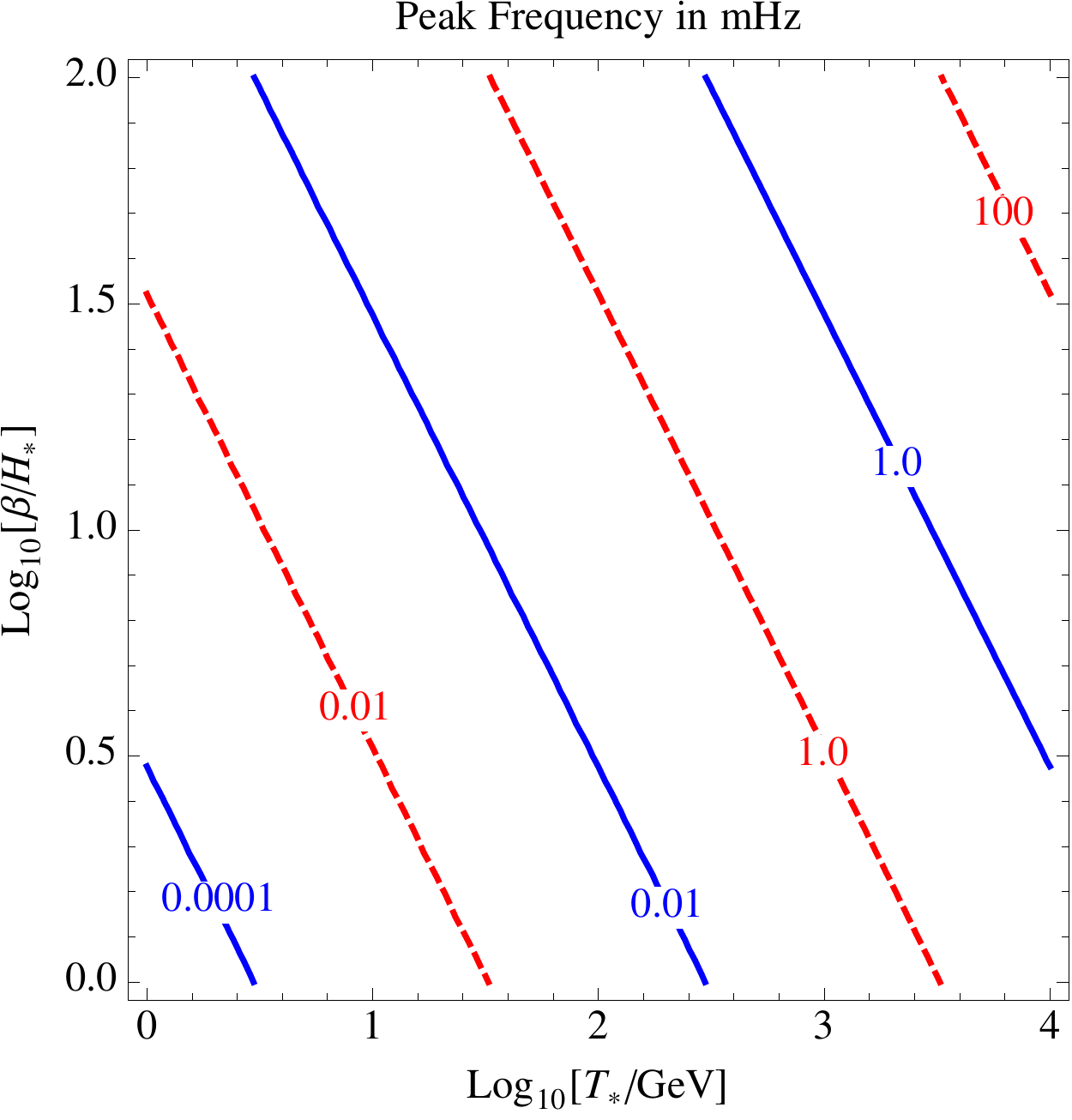}
\hspace*{0.5cm}
\includegraphics[width=0.47\textwidth]{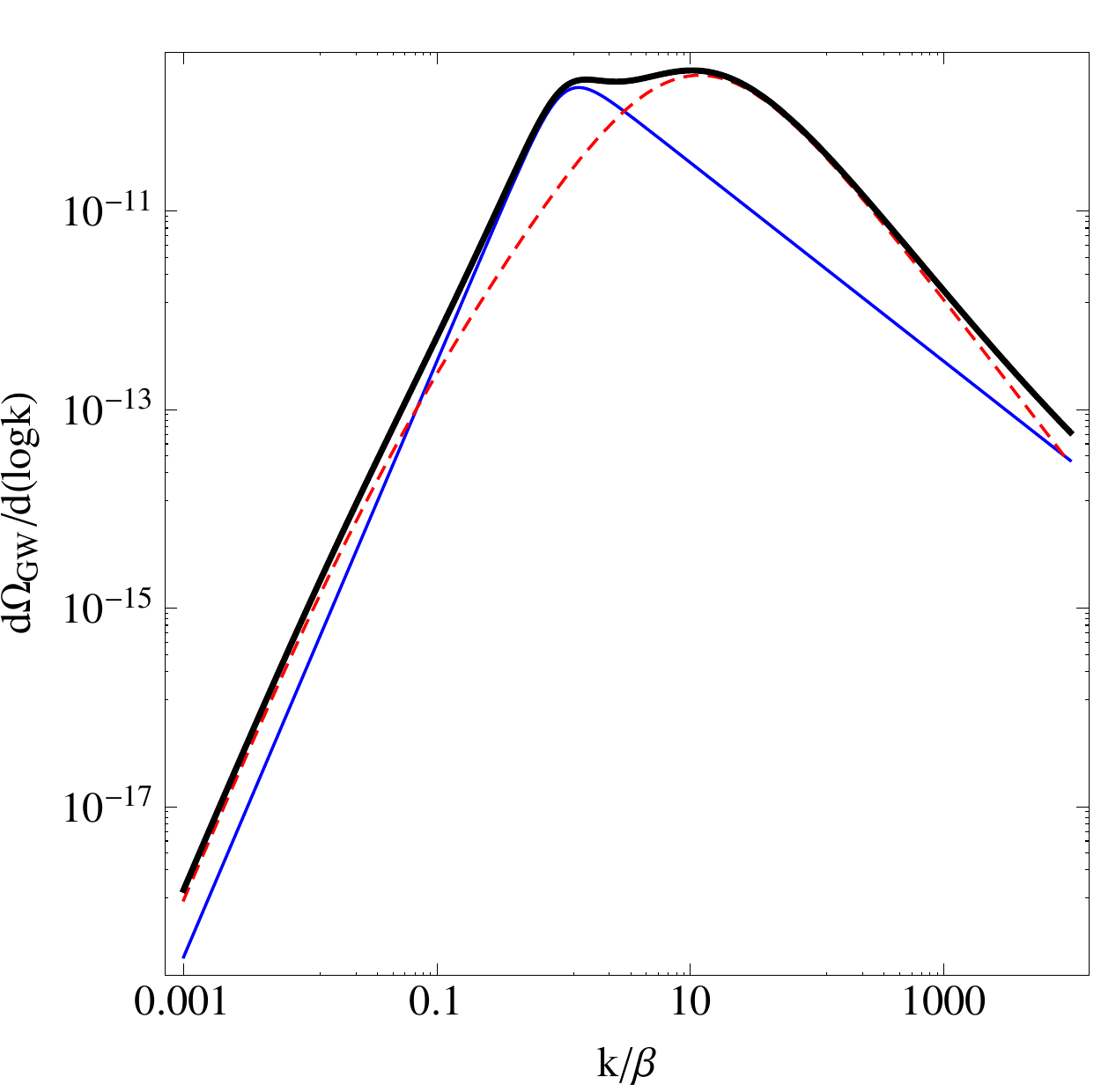}
\caption{Left: Peak frequencies of the GW spectra (in mHz) from bubble collisions (blue,solid) and MHD turbulence (red, dashed) in the $T_* - \beta $ plane, for $v=1.0$. Right: GW spectrum from bubble collisions (blue, solid) and turbulence (red, dashed) as well as the combined spectrum (black, thick), as a function of conformal wave number $k$, for $v=1.0$ and $\beta = 10{\cal H}_*$.  }
\label{fig:peaks}
\end{figure}

Following~\cite{Caprini:2010xv}, we will use $\Omega_{S*} =0.1$ and $\beta = 10 {\cal H}_*$, but $v=1.0$ instead of 0.7. We are now in a position to study the location of the peaks of the GW signals from bubble collisions and MHD turbulence. The bubble collision signal is triangular shaped with a maximum at $k/\beta = \sqrt[4]{3} \approx 1.3$, while the MHD turbulence peaks at somewhat larger wave numbers $k/\beta \approx \pi^2/v$. To obtain physical frequencies, we use Eqn.~(\ref{eqn:f0}) and $f_*/H_* = (\beta/(2 \pi {\cal H})) (k/\beta)$. 
Then the peak locations are
\begin{align}
	f_{\rm peak}^{(B)} & = 3.33\times 10^{-8}~{\rm Hz} \times \left(\frac{g_{*}}{80} \right)^{\frac{1}{6}}  \left(  \frac{T_*}{1~{\rm GeV}}\right)  \left( \frac{\beta}{{\cal H}_*}\right) \,, \quad f_{\rm peak}^{(MHD)} \approx 10 f_{\rm peak}^{(B)} \,.
\end{align}

In Fig.~\ref{fig:peaks} we show the location of the frequency peaks as function of the PT temperature $T_*$ and $\beta$. As expected from Eqn.~(\ref{eqn:f0}), the peak frequencies increase linearly with the transition temperature $T_*$ and with $\beta/{\cal H}_*$. 

The source term $\Omega_{S*}$ can be different for bubble collision and turbulence. Here we will assume that equal amounts of energy act as source for $\Omega^{(B)}_{\rm GW}$ and $\Omega^{(MHD)}_{\rm GW}$. In this case the turbulence signal dominates over the one from bubble collisions over most of the relevant frequency range, see Fig.~\ref{fig:peaks}. The intensity of both signals decreases as $(\beta/{\cal H}_*)^{-2}$, therefore smaller values of $\beta$ are preferable. From Eqn.~(\ref{eqn:MHD}) it might appear that the turbulence signal only decreases as $\beta^{-1}$, however the $k/{\cal H}_*$ term in the denominator gives another power of ${\cal H}_*/\beta$ for $k\gtrsim 1$. 

\

Recent simulations of first order PTs suggest that sound waves generated by the expansion of bubbles could be the dominant source of GWs from these transitions~\cite{Hindmarsh:2013xza,Hindmarsh:2015qta,Kalaydzhyan:2014wca}. Sound waves continue propagating through the early universe after the PT is finished, and decay on a timescale ${\cal H}_*$. Compared to the above discussed spectra, they will therefore not be suppressed as much by the velocity of the transition $\beta$, and the signal could be increased by a factor $(\beta/{\cal H}_*)$ compared to the bubble collision signal, but with a spectrum decaying as $k^{-3}$. This could potentially boost the signal, in particular for cases where the PT is fast, i.e. $\beta/{\cal H} \gg 1$.

\section{Detectability}\label{sec:detect}

In the previous section, we have seen that the peak frequencies of GW signals from GeV-TeV scale PTs are of order $(10^{-6}-10^{-3})$~Hz. Furthermore it is important to note that a broad spectral region around the peak is populated by GWs, from $(10^{-10}-1)$~Hz. 

\begin{figure}
\centering
\includegraphics[width=0.75\textwidth]{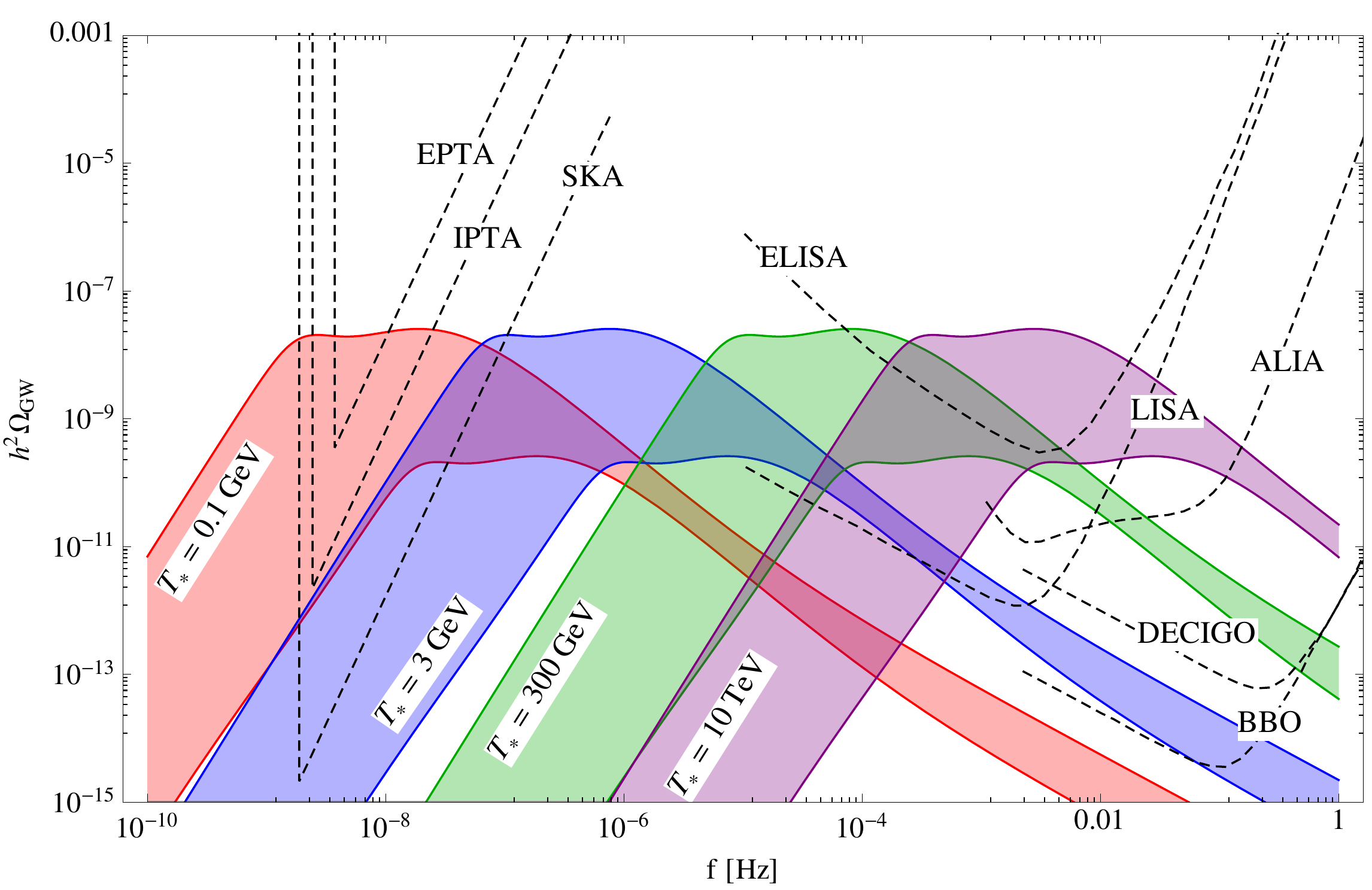}
\caption{GW spectra $\Omega(f)h^2$ for $T_*=0.1$~GeV (SIMP), $T_*=3$~GeV (CDM1, TH models), $T_*=300$~GeV and $T_*=10$~TeV (CDM2 models). The upper (lower) edges of the contours correspond to $\beta={\cal H}$ $(\beta = 10{\cal H})$, and furthermore $v=1$ and $\Omega_{S*}=0.1$ for all curves. The red band $T_*=0.1$~GeV indicates where a signal of the QCD PT would lie if it was strong. 
The projected reach of several planned GW detection experiments is shown (dashed). }
\label{fig:detect}
\end{figure}

GWs with frequencies down to $10^{-5}$~Hz can be probed by satellite based experiments like eLISA~\cite{Seoane:2013qna}, however the sensitivity quickly degrades below $10^{-3}$~Hz. On the other end of the spectrum, pulsar timing arrays (PTA) can probe frequencies in the $(10^{-9}-10^{-7})$~Hz range. In Fig.~\ref{fig:detect} we overlay the expected GW signal for different model parameters with the expected sensitivities of current and planned GW detection experiments (based on~\cite{Moore:2014lga}). 

%

Clearly the most promising signals are those from models with a PT temperature in the TeV range, where the peak region of the GW spectrum falls right into the most sensitive frequency range of satellite experiments. Here the signal should even be detectable for choices of the parameters that are less optimistic than those used for Fig.~\ref{fig:detect}. Models of the CDM2 type naturally fall into this region, but also the CDM1 models can be viable with a confinement scale in that region. 

The fact that TeV dark sectors predict an observable GW signal is not surprising, since here we are in the energy range of the electroweak PT or beyond, and the observability in particular of TeV scale strong PTs has been noted before~\cite{Randall:2006py,Konstandin:2010cd,Konstandin:2011dr}. The novelty here is that the dynamics leading to this strong PT does not have to be connected to the electroweak sector of the SM, and is therefore not in tension with the non-observation of new physics at the LHC. 

For models with $T_* \sim (1-10)$~GeV the situation is a bit more difficult, since the signal peak ends up in a frequency region where neither PTA nor (e)LISA are sensitive. Looking at the $T_*=3$~GeV curve in Fig.~\ref{fig:detect} more closely, we see that in the best case scenario, for $\beta={\cal H}$, both PTAs and LISA would be able to detect parts of the GW spectrum. For larger $\beta$ the signal quickly drops out of the PTA sensitivity region, however LISA remains sensitive. This is due to the increase of the observed frequency with $(\beta/{\cal H})$, which partially compensates the overall $({\cal H}/\beta)^2$ drop of the signal in the LISA sensitivity region. Therefore there is a chance to detect a GW signal from the CDM1 and TH models, even if the predicted frequency range is not optimal. 

%
%
%

Recently the EPTA experiment has reported the first limit on a stochastic GW background~\cite{Lentati:2015tma}, which probes values of $\Omega_{\rm GW}h^2$ of order $10^{-8}$ in the ($10^{-9}-10^{-7}$)~Hz range. Their limit can not directly be displayed in Fig.~\ref{fig:detect} since it depends on the assumed spectrum, but it is nevertheless interesting to note that GWs from a 100~MeV scale dark sector could already be detectable. 

Another way of interpreting these results is that GW searches have the potential to observe signals of completely unknown physics in the early universe, while the absence of a signal in a certain frequency range would place (very weak) bounds on the dynamics of new physics at the corresponding energy scale.

\section{Complementarity}\label{sec:compl}

Signals of the new physics models introduced in Sec.~\ref{sec:models} are being searched for at collider experiments and in direct and indirect dark matter searches. The possibility to observe such a signal in those experiments always relies on sufficiently strong non-gravitational interactions of the dark sector with the SM. Instead the GW signal is unique in the sense that it probes the gravitational effects of a model at very early times, and, while the detectability of a signal eventually still depends on model parameters, those are relatively independent of whether the model is detectable at colliders or in dark matter searches.  

In CDM1 type models, or more general, hidden valley models with a new strong interaction, the dark sector is neutral under all SM interactions, and communicates only through a heavy mediator. If that mediator is in the TeV range, both direct detection experiments and the LHC could discover these models, but for larger masses it becomes increasingly difficult. From the DM perspective on the other hand, mediator masses of 10s of TeV are acceptable, such that part of the parameter space will remain unexplored in the near future. The GW signal instead is completely independent of the mediator mass, such that it could be detected even if the model was not discovered before, and thanks to the straightforward connection of mass scales and GW frequencies, could even motivate future experimental efforts.

The situation is similar for TH models, which mainly communicate with the SM through the Higgs portal. Here the parameter that controls detectability is not a mass scale but the smallness of the mixing angle of the Higgs with its twin partner. First hints for such a model could come from deviations in Higgs couplings and from exotic Higgs decays, but even then it would be difficult to uncover the whole structure. A GW signal in the right energy range could provide much information about the dynamics of these models near their confinement scale. 

Finally the CDM2 type models have many detectable features at hadron colliders. However the upper limit on their mass scale is of order 100~TeV, which comes from the requirement that dark matter is not overproduced. Such high scales are not in reach of the current or next generation of collider experiments, but seem very accessible by GW searches, since the signal moves into the most sensitive frequency region of LISA and other satellite based experiments. 

Overall we see that GW experiments provide a unique window to explore the dynamics of these models in the early universe, even if they are not discovered in the near future. 
Of course it would be even more interesting if such a dark sector was discovered at the LHC. The GW signal would then provide an important and unique probe of the physics of those models in the early universe.

Moving beyond the concrete models discussed above, perturbative unitarity constrains the mass of thermal DM to be below 110~TeV~\cite{Griest:1989wd,Blum:2014dca}.  For composite non-perturbative DM this limit does not apply directly\footnote{We thank T.~Cohen for discussion of this point.}, instead a lower bound on the radius of the extended object can be obtained, $R \lesssim (100~{\rm TeV})^{-1}$. It is reasonable to expect the radius $R$ to be of order of the inverse mass, which again implies an upper bound on the DM mass of order 100~TeV. GW signals could therefore be a unique probe of the thermal DM paradigm.

\section{Conclusions}

Models beyond the standard model with a confining dark sector can lead to unexpected phenomenological signatures. Here we have explored the possibility to detect gravitational waves due to a first order phase transition at the confinement scale $\Lambda_d$. The main messages from this paper are as follows:
\begin{itemize}
	\item Different from QCD, dark sectors with QCD-like interactions can undergo strong first order phase transitions, with only mild constraints on the particle content of the theories. 
	\item Several classes of new physics models that are currently being explored fulfil the criteria for first order PTs. The physics problems these models are trying to address, either dark matter or naturalness, constrain the confinement scales and therefore the temperature range of the phase transition. 
	\item The GW signals originating from these dark phase transitions are in the detectable frequency range of future GW experiments, either at (E)LISA and BBO for high scale models, or in PTA experiments for the lower end of the spectrum. 
\end{itemize}
Depending on other aspects of the model, GW signals will either provide complementary information about the models in question, or might even be the the best option to find evidence for these models of new physics. Different from the electroweak PT, a null result at LHC will not strongly disfavour a strong PT in a dark sector, although of course a confirmation of their existence would be more exciting. 

A shortcoming of the present work is a lack of precise quantitative predictions. The bubble velocity $v$ as well as the time scale of the phase transition $\beta^{-1}$ and the energy fraction $\Omega_{S*}$ are currently unknown, and are set to optimistic (but not unrealistic) values. Two approaches seem possible to improve upon this situation: On one side, lattice simulations could be used to measure quantities like the latent heat and the surface tension, which are related to the above parameters and can be used to obtain a more quantitative prediction for the GW spectra. Alternatively, one could attempt to construct a holographic dual for some of these theories, and analyse the PT in that setup.

\paragraph{Acknowledgements} I would like to thank M.~Laine for very valuable discussions regarding the order of the PT and for encouragement and comments on the manuscript. Furthermore I would like to thank A.~Kurkela, T.~Konstandin, M.~Mccullough, D.~Stolarski and A.~Weiler for useful discussions and comments on the manuscript, and the organisers and participants of the eLISA and Neutral Naturalness workshops at CERN for providing a stimulating environment to finalise this work.

 \end{document}